\documentclass[aps,prl,twocolumn,groupedaddress,showpacs]{revtex4}
\usepackage{epsfig}

\begin{document}

\title{Effective-range approach and scaling laws for electromagnetic strength
in neutron-halo nuclei}

\author{S. Typel}
\affiliation{Gesellschaft f\"{u}r Schwerionenforschung mbH (GSI),
Planckstra\ss{}e 1, D-64291 Darmstadt, Germany}

\author{G. Baur}
\affiliation{Institut f\"{u}r Kernphysik, Forschungszentrum J\"{u}lich,
D-52425 J\"{u}lich, Germany}

\date{\today}

\begin{abstract}
We study low-lying multipole 
strength in neutron-halo nuclei.
The strength depends only on a few low-energy constants:
the neutron separation energy,
the asymptotic normalization coefficient of the bound state
wave function, and the scattering length that contains
the information on the interaction in the continuum.
The shape of the transition probability 
shows a characteristic dependence on few
scaling parameters and the angular momenta.
The total $E1$ strength is related to the root-mean-square radius of the 
neutron wave function in the ground state and shows corresponding
scaling properties.
We apply our approach to the $E1$ strength distribution of ${}^{11}$Be.
\end{abstract}


\maketitle

Electromagnetic dipole strength in stable nuclei 
is concentrated in the giant dipole resonance.
In recent years it was found  that there is an appreciable 
low-lying $E1$ strength in light neutron-rich nuclei
\cite{Lei01,Jon04}.
This phenomenon is especially pronounced in 
neutron-halo nuclei, where it can be explained essentially
as a single-particle effect \cite{Han87,Jen04}.
Examples are $^{11}$Be \cite{Nak94,Pal03}
and $^{19}$C \cite{Nak99}.
This low-lying strength was observed
mainly in Coulomb dissociation experiments, 
for a recent review see  \cite{Bau03}.

The dissociation cross section  directly reflects 
the dipole strength 
distribution or reduced transition probability $dB(E1)/dE$.
Photodissociation
of and radiative capture leading to halo nuclei
provide equivalent information.  
The measured strength distributions show simple features.
They tend to be universal
when plotted in the appropriate reduced parameters
\cite{Typ01a}
because low-energy processes do not depend on certain details 
of the interaction. Consequently,
effective field theories are nowadays used for 
the description of halo nuclei
\cite{Ber02}.
In Ref.\ \cite{Bir99} nonrelativistic 
two-body scattering by a short range potential was studied using the 
renormalisation group. It was found that the expansion around 
the nontrivial fixed point is equivalent to the effective-range 
(ER) expansion. 

It is the purpose of this Letter to apply the 
ER approach to one-neutron halo nuclei and 
find scaling laws for the transition strength.
This work is similar in spirit to effective field theories:
the effects of unknown short distance behaviour are
parametrized by a few low-energy constants. 
It is  not our aim  to relate these parameters to the many-body
physics.

General expressions for electromagnetic strength distributions
in halo nuclei were derived before (see, e.g.,
\cite{Muk02} and references therein for proton+core systems), however, 
often neglecting the nuclear continuum interaction. Also
expansions for small relative energies were obtained \cite{Bay04}.
The application of ER theory 
to the study of halo nuclei was considered 
before, e.g., in the description of ground state properties  
\cite{Esb97}, and  for radiative capture cross sections into
$s$-, $p$- and  $d$-wave bound states
taking into account the interaction only for $s$-waves
assuming a zero-range potential \cite{Kal96}.
The effect of the interaction in the continuum states
on direct neutron capture and photodisintegration of $^{13}$C 
was studied in \cite{Men95} and \cite{Ots94}.
While they find a sensitivity on neutron
optical model parameters for $s \rightarrow p$ capture, this sensitivity
is strongly reduced for the cases of $p \rightarrow s$ 
and $p \rightarrow d$ capture.

The prototype of a neutron halo nucleus is the deuteron.
Radiative capture or photodissociation at low energies 
are well described by the binding energy and the effective range,
see e.g.\ \cite{Bla79}.
Quite similar,
the low-lying $E1$ strength in neutron-halo nuclei
is determined by 
two parameters if the continuum interaction is neglected: 
one is the binding energy (or neutron separation energy $S_{n}$) 
which determines the
overall shape of the $dB(E1)/dE$ distribution, the other one is 
a normalization constant.  
In the analogous case of $^{19}$C, the binding energy
could be determined directly from the shape of the 
observed 
strength distribution \cite{Nak99}. 
Effects of the neutron-core interaction are parametrized efficiently
by the ER approximation. 

The cross section for photodissociation of a nucleus $a$
into a neutron $n$ and a core $c$ is given by
\begin{eqnarray} \label{eq:sabs}
 \lefteqn{\sigma_{\pi \lambda}(a+\gamma \to n+c) =}
 \\ & &  \nonumber
 \frac{\lambda+1}{\lambda} \frac{(2\pi)^{3}}{[(2\lambda+1)!!]^{2}}
 \left(\frac{E_{\gamma}}{\hbar c}\right)^{2\lambda-1}
 \frac{dB(\pi \lambda)}{dE} \: .
\end{eqnarray}
The reduced transition probability for electromagnetic 
multipolarity $\pi \lambda$ is denoted by $dB(\pi \lambda)/
dE$. The photon energy is given by $E_{\gamma}=E+S_{n}$
with $S_{n}=\hbar^{2}q^{2}/(2\mu)$ where
$q$ is the inverse
decay length of the bound state wave function and
$\mu$ is the reduced mass.
Using detailed balance, the corresponding radiative
capture cross section can be obtained from this expression.
With the equivalent photon number, cross section
(\ref{eq:sabs}) determines 
the (first order) Coulomb dissociation cross section
\cite{Bau03}.

The strength distribution $dB(E\lambda, l_{i} \to l_{f})/ dE$
for a certain transition 
depends on the corresponding matrix element
that contains a radial integral with the wave functions 
of the initial and final states
with orbital angular momenta $l_{i}$
and $l_{f}$, respectively.
(We consider spinless
neutrons and cores, a generalization is obvious.)
At low energies $E=\hbar^{2}k^{2}/(2\mu)$, 
the main contribution to the radial integral
arises from radii larger than the radius $R$ of the nucleus.
This will be true for all possible values 
of the angular momenta involved.
In the figures of Refs.\ \cite{Men95,Ots94} 
it can very well be seen that the radial integrals 
are dominated by the outside region. 
Away from a resonance,
a case which we always assume here, the continuum wave function is 
small inside the nuclear radius. In a hard sphere model
it is exactly zero. 

For the neutron+core
case the dimensionless reduced radial
integral is well approximated by
\begin{eqnarray} \label{eq:idef}
 {\mathcal I}_{l_{i}}^{l_{f}}(\lambda) 
 & = & - i^{l_{i}} k q^{\lambda+2} 
 \int_{R}^{\infty} dr \: r^{\lambda+2} 
 \:  h^{(1)}_{l_{i}} (iqr) \\
 \nonumber & & \times
  \left[ \cos(\delta_{l_{f}})j_{l_{f}}(kr)- 
 \sin(\delta_{l_{f}})y_{l_{f}}(kr)   \right]
\end{eqnarray}
with spherical Bessel ($j_{l}$), Neumann ($y_{l}$), 
and Hankel ($h^{(1)}_{l}$) functions, respectively,
that describe the behaviour of the radial wave functions
beyond the range of the nuclear potential.
The phase shift $\delta_{l_{f}}$ contains the information
on the interaction in the final continuum state.
Introducing the dimensionless shape functions
\begin{equation} \label{eq:sdef}
 {\mathcal S}_{l_{i}}^{l_{f}}(\lambda)
 =  \frac{q}{k} \left| 
 {\mathcal I}_{l_{i}}^{l_{f}}(\lambda)
\right|^{2}
\end{equation}
the reduced transition probability is given by
\begin{eqnarray} \label{eq:dbelde}
 \lefteqn{\frac{dB}{dE} (E\lambda, l_{i}  \to l_{f})=
  \left[Z_{\rm eff}^{(\lambda)}e\right]^{2} 
 \frac{2\mu}{\pi\hbar^{2}}}
 \\ \nonumber & & \times
 \frac{2\lambda+1}{4\pi}
 ( l_{i} \: 0 \: \lambda \: 0 | l_{f} \: 0 )^{2}
  \frac{\left|C_{l_{i}}\right|^{2}}{q^{2\lambda+3}}
 {\mathcal S}_{l_{i}}^{l_{f}}(\lambda)
\end{eqnarray}
with the effective charge number $Z^{(\lambda)}_{\rm eff}
=Z_{c}[m_{n}/(m_{n}+m_{c})]^{\lambda}$
and the asymptotic normalization constant (ANC) $C_{l_{i}}$ of the
bound state wave function.
For  neutron+core systems the dipole 
transitions are dominant since
$E2$ transitions are suppressed by the small effective charge.
 
The ANC of the actual many-body wave function as it appears in 
(\ref{eq:dbelde})
consists of the ANC of a neutron single-particle
wave function and a spectroscopic factor which 
accounts for the many-body aspects.  The single-particle ANC
is determined by the normalization of the wave function.
It depends on $q$ with the scaling behaviour
$|C_{l_{i}}|^{2} \propto q$ for $l_{i}=0$ and 
$|C_{l_{i}}|^{2} \propto q \gamma^{2l_{i}-1}$ for $l_{i}>0$
as seen, e.g., in a square-well model.
Here, we have introduced the dimensionless scaling parameter
$\gamma=qR$.
The factor $\left|C_{l_{i}}\right|^{2}/q^{2\lambda+3}$ in the
reduced transition probability (\ref{eq:dbelde}) depends only on
properties of the ground state. It directly shows the
scaling with the characteristic parameter $q$. 
Clearly, for small
separation energies of the neutron a large transition strength can be
expected.

The reduced radial integrals (\ref{eq:idef}) have the general form 
\begin{eqnarray} \label{eq:iredgen}
 \lefteqn{{\mathcal I}_{l_{i}}^{l_{f}}(\lambda) =
 \frac{\gamma \exp(-\gamma)}{(\gamma^{2}+\kappa^{2})^{\lambda+1}} 
 \left(\frac{\gamma}{\kappa}\right)^{l_{f}}}
 \\ \nonumber & & \times
 \left[ {\mathcal R}_{l_{i}}^{(+)l_{f}}(\lambda) \cos (\kappa+\delta_{l_{f}}) 
 + {\mathcal R}_{l_{i}}^{(-)l_{f}}(\lambda) \sin(\kappa+\delta_{l_{f}})\right]
\end{eqnarray}
with $\kappa = kR$
where the rational functions ${\mathcal R}_{l_{i}}^{(\pm)l_{f}}(\lambda)$ 
can be found for all relevant  values of $l_{i}$ and $l_{f}$
by means of the recursion relations
\begin{eqnarray}
 \lefteqn{{\mathcal R}_{l_{i}}^{(\pm)l_{f}+1}(\lambda+1) =
  \left[ 2\kappa^{2}(\lambda+1) \right.}
 \\ \nonumber  & & \left.
 + (\gamma^{2}+\kappa^{2})(2l_{f}+1)\right]
 {\mathcal R}_{l_{i}}^{(\pm)l_{f}}(\lambda)
 \\ \nonumber & & 
 - \kappa (\gamma^{2}+\kappa^{2}) \left[
 \frac{d}{d\kappa} {\mathcal R}_{l_{i}}^{(\pm)l_{f}}(\lambda)
 \pm {\mathcal R}_{l_{i}}^{(\mp)l_{f}}(\lambda)\right] \: ,
 \\
 \lefteqn{{\mathcal R}_{l_{i}+1}^{(\pm)l_{f}}(\lambda+1) =
 \left[ 2\gamma^{2}(\lambda+1) \right.}
 \\ \nonumber & & \left.
 + (\gamma^{2}+\kappa^{2})(l_{i}+\lambda+1-l_{f}+\gamma)\right]
 {\mathcal R}_{l_{i}}^{(\pm)l_{f}}(\lambda)
 \\ \nonumber & & 
 - \gamma (\gamma^{2}+\kappa^{2}) 
 \frac{d}{d\gamma} {\mathcal R}_{l_{i}}^{(\pm)l_{f}}(\lambda) 
\end{eqnarray}
starting from
\begin{equation}
 \begin{array}{ll}
 {\mathcal R}_{0}^{(+)0}(0) = \kappa  \: , & 
 {\mathcal R}_{0}^{(-)0}(0) = \gamma \: ,
 \\ & \\ 
 {\mathcal R}_{1}^{(+)1}(0) = - \kappa \: , & 
 {\mathcal R}_{1}^{(-)1}(0) = 
 [\kappa^{2}(1+\gamma)+\gamma^{2}]/\gamma^{2}  \: .
 \end{array}
\end{equation}
The radial integral involving the regular scattering 
wave function shows a $r^{\lambda-l_i+l_f+1}$ behavior
for small values of $r$,
thus it can be extended to $R=0$. On the other hand,
the radial integral involving the irregular function
shows a $r^{\lambda-l_i-l_f}$ behaviour. Thus for $\lambda=1$
we can extend the integral to zero only for $l_i=0,l_f=1$
and $l_i=1, l_f=0$ and have convergence. In the other cases,
one has to cut off the integral for a finite $R$ value.

For low energies we can use the 
effective-range approximation 
$ k^{2l+1} \cot(\delta_{l}) = - 1/a_{l} + r_{l} k^{2}/2 + \dots$
for the phase shifts $\delta_{l}$
with the scattering length $a_{l}$ and the effective range $r_{l}$
in order to take the interaction in the continuum state into account.
(Note that $a_{l}$ and $r_{l}$ have the dimension of a
length only for $l=0$.)
We restrict ourselves to the lowest order term, i.e.\
\begin{equation} \label{rsl}
 \tan(\delta_{l}) = - (xc_{l}\gamma)^{2l+1}
\end{equation}
with the dimensionless reduced scattering length $c_{l}$ defined by
$c_{l}^{2l+1} = a_{l}/R^{2l+1}$
and the ratio  $x=\kappa/\gamma=k/q=\sqrt{E/S_{n}}$.
We assume that this is a reasonable approximation
of the phase shift.  We further assume that
the scattering length has a ``natural'' value with 
$c_{l}$ being $O(1)$. For example, for $s$-wave scattering from a hard sphere
with radius $R$ we have $c_{0}=1$. 
In exceptional cases, 
this assumption is not fulfilled, e.g.\
if there are resonances in the low-lying continuum or
subthreshold states. For an $s$-wave halo nucleus bound
by a zero-range force the scattering length is given by 
$a_{0} = 1/q$, i.e.\ $c_{0} = 1/\gamma $. In such a
case the scattering length in the $s$-wave is unnaturally large; it 
diverges in the halo limit $\gamma \rightarrow 0$.

We can expand the analytical results for $S_{l_{i}}^{l_{f}}(\lambda)$ 
in terms of the small
parameter $\gamma$ and obtain for the most important cases for dipole
transitions
\begin{eqnarray}
 \label{eq:s011}
 {\mathcal S}_{0}^{1}(1) & = &
 \frac{4x^{3}}{(1+x^{2})^{4}}
 \left[1 - c_{1}^{3}(1+3x^{2})\gamma^{3} + \dots \right]
 \: ,
 \\
 {\mathcal S}_{1}^{0}(1) & = & 
 \frac{x(3+x^{2})^{2}}{(1+x^{2})^{4}} 
 \left[ 1 - \frac{4c_{0}}{3+x^{2}} \gamma + \dots \right]
 \: .
\end{eqnarray}
The term in front of the square parenthesis in Eq.\ (\ref{eq:s011})
is the well-known result without
continuum interaction, see, e.g., \cite{Nak94,Ots94}.
In general there will be a modification of the shape function.
It is getting more and more important
for larger binding energies of the halo nucleus, i.e.\ larger $\gamma$. 
Transitions with $l_f=l_i-1$ are more affected by this final state 
interaction than transitions with $l_f=l_i+1$.
In the case $s \to p$ the first correction
to the shape function for $\gamma \to 0$ appears only in the 
$\gamma^{3}$ term proportional to $c_{1}^{3}$. There are no corrections
of lower order in $\gamma$ independent of the final state interaction.
This explains the remarkable uniformity of the characteristic shape
for the low-lying $dB(E1)/dE$ strength in halo nuclei like ${}^{11}$Be
and ${}^{19}$C. 
For the $p \to s$ transition there is a greater sensitivity to the 
final-state interaction. It appears already as a correction linear in
$\gamma$ and $c_{0}$. This is in accord with the results of Ref.\ 
\cite{Men95}.

The total transition strength can be obtained by integrating
(\ref{eq:dbelde}) over the relative energy $E$. It is given by
\begin{eqnarray} \label{eq:bel}
 \lefteqn{B (E\lambda, l_{i}  \to l_{f})  = 
   \left[Z_{\rm eff}^{(\lambda)}e\right]^{2}}
 \\ \nonumber & & \times
 \frac{2\lambda+1}{4\pi}
 ( l_{i} \: 0 \: \lambda \: 0 | l_{f} \: 0 )^{2}
   \frac{\left|C_{l_{i}}\right|^{2}}{q^{2\lambda+1}}
 {\mathcal T}_{l_{i}}^{l_{f}}(\lambda) 
\end{eqnarray}
with
\begin{equation} \label{eq:tdef}
{\mathcal T}_{l_{i}}^{l_{f}}(\lambda)
 = \frac{2}{\pi} \int_{0}^{\infty} dx \: x \:
{\mathcal S}_{l_{i}}^{l_{f}}(\lambda) 
 =  \frac{2}{\pi} \int_{0}^{\infty} dx \:
 \left| {\mathcal I}_{l_{i}}^{l_{f}}(\lambda) \right|^{2} \: .
\end{equation}
For dipole transitions and no final-state interaction (i.e.\ $c_{l}=0$
corresponding to a plane wave in the final state) we find
\begin{eqnarray} \label{eq:gent} 
 & &  {\mathcal T}_{1}^{0}(1) = {\mathcal T}_{1}^{2}(1) =
 \frac{1}{4}(5+6\gamma+2\gamma^{2}) e^{-2\gamma} \: ,
 \\ & & {\mathcal T}_{0}^{1}(1) = \frac{1}{4}
(1+2\gamma+2\gamma^{2})e^{-2\gamma} \,
 \\  & & {\mathcal T}_{2}^{1}(1)
  = \frac{1}{4\gamma}  (36+37\gamma+14\gamma^{2}+2\gamma^{3})
 e^{-2\gamma} 
 \: .
\end{eqnarray}
A finite value of $\gamma$ leads to a reduction of the total strength
in the continuum as compared to the extreme halo limit $\gamma=0$.
With final-state interaction,  the functions
${\mathcal T}_{l_{i}}^{l_{f}}(\lambda)$ also
depend on the reduced scattering length $c_{l}$, however rather lengthy
expressions are obtained.

The value of $B(E1, l_{i}) = \sum_{l_{f}}B(E1, l_{i} \to l_{f})$ 
for all $E1$ transitions from a state with orbital angular momentum $l_{i}$
is directly related by the non energy-weighted sum rule
\begin{equation} \label{eq:sr}
 B(E1, l_{i}) =
 \left[ Z_{\rm eff}^{(1)} e \right]^{2}
 \frac{3}{4\pi} \langle r^{2} \rangle_{l_{i}} 
\end{equation}
to the root-mean-square radius
$\langle r^{2} \rangle_{l_{i}}$ of the same bound 
state, see, e.g., \cite{Jon04}.
The rms radius scales with $\gamma$ for different values of the bound-state
orbital angular momentum $l_{i}$ as
$\langle r^{2} \rangle_{0} \propto R^{2}/\gamma^{2}$,
$\langle r^{2} \rangle_{1} \propto R^{2}/\gamma$, and
$\langle r^{2} \rangle_{l_{i}} \propto R^{2}$
for $l_{i}\geq 2$ \cite{Jen04}. 
On the other hand, the energy-weighted sum rule (Thomas-Reiche-Kuhn sum rule)
gives a constant independent of 
$l_{i}$. From the above we see clearly that the low-lying strength is
most pronounced for $s$-wave bound states and to a lesser degree for
$p$-wave bound states in halo nuclei.
Considering the $q$ dependence of $B(E1, l_{i} \to l_{f})$
in (\ref{eq:bel})
the scaling of $\langle r^{2} \rangle_{l_{i}}$ with $\gamma$
is exactly reproduced. With no final-state interaction the full
transition strength as predicted by the sum rule is found in the continuum.
However, if there are bound states in the particular final state the continuum
contribution to the total strength is reduced and the shape will
be distorted. This will correspond to 
a non-zero value of $c_{l}$.

\begin{table}
\caption{\label{tab:cl}
Dimensionless quantities $c_{1}^{j}$, see Eq.\ (\ref{rsl}), and
spectroscopic factors $C^{2}S$ from the fit
to experimental data \cite{Pal03} of ${}^{11}$Be Coulomb dissociation.} 
\begin{ruledtabular}
\begin{tabular}{cccc}
 model & $c_{1}^{3/2}$ & $c_{1}^{1/2}$ & $C^{2}S$ \\
 \hline
 effective-range  & $-0.41(86,-20)$ & $2.77(13,-14)$ & $0.704(15)$ \\
 Woods-Saxon  & $-0.46(70,-14)$\footnote{At energy $E=0.4$~MeV in the peak
 of the strength distribution.} 
 & $1.87(3)^{a}$ & $0.696(15)$\\
\end{tabular}
\end{ruledtabular}
\end{table}

\begin{figure}[t]
\epsfig{file=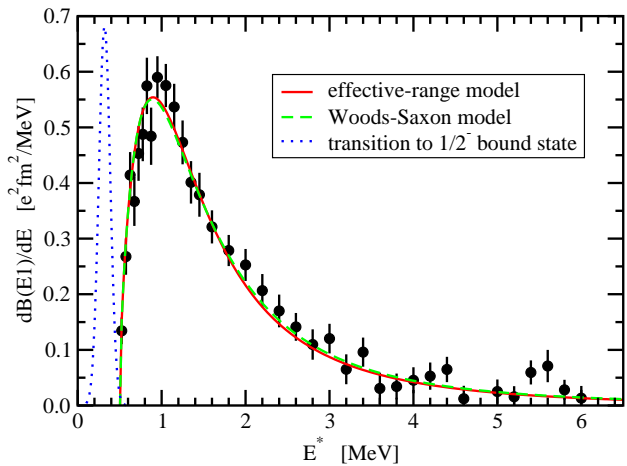,width=8.5cm} 
\caption{Reduced transition probability as a function of the
excitation energy $E^{\ast}=E+S_{n}$
compared to experimental data 
extracted from the Coulomb breakup of ${}^{11}$Be \cite{Pal03}.}
\label{fig:dbde}
\end{figure}

Let us now apply the general theory to $^{11}$Be, which 
shows a very pronounced single particle halo structure
with a neutron bound by 504~keV in the $\frac{1}{2}^{+}$ ground state
that can be considered as a $2s_{1/2}$ state.
Assuming a radius of $R=2.78$~fm the neutron separation energy
corresponds to a small value $\gamma = 0.4132<1$ typical for a halo nucleus.
Unlike the case of the deuteron, where there is no
excited bound state, there is a $\frac{1}{2}^{-}$-bound state at 320~keV
excitation energy.
We can regard it to a good approximation as a $1p_{1/2}$
single-particle state. 
We perform a $\chi^{2}$ fit of our analytical ER approach
(not using the expansion for small $\gamma$) to the experimental 
Coulomb dissociation data \cite{Pal03} 
by adjusting the reduced scattering lengths $c_{1}^{3/2}$
and $c_{1}^{1/2}$ in the two $p$-wave channels 
(see Table \ref{tab:cl}) and the ground state ANC $C_{0}$.
The quantity $c_{1}^{1/2}$ is definitely unnaturally large
as expected from the existence of the weakly bound $\frac{1}{2}^{-}$ state.
In Figure \ref{fig:dbde} our model fit (solid line) is compared to
the experimental strength distribution 
\cite{Pal03}, including corrections due to the detector response.

The obtained ANC $C_{0} = 0.724(8)$~fm$^{-1/2}$
can be translated into a spectroscopic factor $C^{2}S$ by comparing
to the ANC of a normalized single-particle wave function generated from a
Woods-Saxon (WS) 
potential with a depth adjusted to the correct neutron separation
energy. Assuming a radius of $R=2.78$~fm as above 
and a diffuseness parameter of $a=0.65$~fm we obtain $C^{2}S=0.704(15)$ 
which compares well
with the spectroscopic factors from experiment and theory as given
in Ref.\ \cite{Pal03}. The model ground state wave function has a
rms radius of $7.172$~fm that corresponds to a total
strength of $B(E1)=1.645$~e${^2}$fm${}^{2}$
according to (\ref{eq:sr}). Applying the spectroscopic factor we obtain
a total strength of $1.159(36)$~e${^2}$fm${}^{2}$.
The experimental continuum strength 
\cite{Pal03} exhausts about $78(5)$~\% of the sum rule. 
The ground state transition probability to the first excited state 
is experimentally known to be 
$B(E1) = 0.116(12)$~e${}^{2}$fm${}^{2}$ \cite{Gla98}
corresponding to $10(1)$~\% of the sum rule.
This bound state transition strength 
is indicated in Figure \ref{fig:dbde} as a dotted line assuming an
arbitrary FWHM of 0.160~MeV.
The remaining $12(6)$~\% can be attributed to the
$p_{3/2}$ state that is occupied.
Comparing the experimental transition strength to results
of a plane-wave calculation would underestimate the extracted
spectroscopic factor. 

Our ER approach can be compared to an
exact calculation of the matrix elements with wave functions 
obtained from WS potentials. We assume the same shape
of the potential with $R$ and $a$ as given above. 
Repeating the $\chi^{2}$ fit by adjusting
the depth of the potential in the two $p$ waves independently
we find $C^{2}S = 0.696(15)$ 
close to the result of our ER approach.
The exact calculation (dashed line in Figure \ref{fig:dbde})
almost coincides with our ER approximation.
We can also extract the dimensionless quantities $c_{1}^{j}$
according to Eq.\ (\ref{rsl}).  
Their modulus decreases with
relative energy (much more strongly for $c_{1}^{1/2}$ than for
$c_{1}^{3/2}$). Since the $c_{1}^{j}$
in our ER approach are constant they can be considered as an
average of the actual values 
over the peak of the strength distribution. Taking this into
account a reasonable agreement of the two approaches is found again
(Table \ref{tab:cl}).

In conclusion we find that low-lying 
electric multipole strength in one-neutron halo nuclei
can be described effectively by a few low-energy constants:
the binding energy, the scattering length
and the asymptotic normalization constant.
We apply our ER approach to the halo nucleus ${}^{11}$Be and find
a remarkable agreement with an exact calculation.
The exhaustion of the strength due to the bound state affects
the extraction of the ANC and of the spectroscopic factor. In our approach
there is no need to determine parameters of optical potentials
from elastic scattering.

The present 
theoretical approach will be applied to other 
halo nuclei. It will be extended to higher multipolarities and to
proton halo nuclei in a forthcoming publication \cite{Typ04b}.
An effective-range theory of 
two-neutron halo nuclei like ${}^{11}$Li
would be very interesting but is certainly much more complex \cite{Jen04}. 
An early work related to this problem is \cite{Mig73}.

We are grateful to T. Aumann and H. Emling for useful discussions.

\end{document}